\documentclass{aa_final}

\usepackage{graphicx}
\usepackage[update,prepend]{epstopdf}
\usepackage{natbib}

\def\ik {{\it Kepler}}

\begin{document}

\title{Dearth of short-period Neptunian exoplanets---a desert in period-mass and period-radius planes}
\author{
T. Mazeh, T. Holczer and S. Faigler 
}
\institute{School of Physics and Astronomy, Raymond and Beverly Sackler Faculty of
Exact Sciences, Tel Aviv University, Tel Aviv, Israel\\
\email{mazeh@post.tau.ac.il}
}
\date{Received 31 December 2015 / Accepted 23 February 2016}
\abstract
{A few studies have reported a significant dearth of exoplanets
with Neptune mass and radius with orbital periods below $2$--$4$ d. This cannot be explained by observational biases because many Neptunian planets with longer orbital periods have been detected.
The existence of this desert is similar to the appearance of the so-called brown-dwarf desert that suggests different formation mechanisms of planets and stellar companions with short orbital periods. Similarly,  the Neptunian desert might indicate different mechanisms of formation and evolution for hot Jupiters
and short-period super-Earths.
We here follow a previous study and examine the location and shape of the desert
 in both the period-mass and period-radius planes, 
 using the currently available large samples of planets. 
 The desert in the period-mass plane has a relatively sharp upper edge, with a planetary mass that is inversely proportional to the planetary orbital period, while the lower, somewhat blurred, boundary is located along masses
that are apparently linearly proportional to the period. 
The desert in the period-radius plane of the transiting planets is less clear. 
It seems as if the radius along the upper boundary is inversely proportional to the period to the power of one-third, while the lower boundary shows a radius that is proportional to the period to the power of two-thirds.
The combination of the two upper bounds of the desert, in the period-mass and period-radius planes, yields a planetary mass-radius relation of 
$R_{\rm p}/R_{\rm Jup}\simeq (1.2\pm0.3)(M_{\rm p}/M_{\rm Jup})^{0.27\pm0.11}$
 for $ 0.1\lesssim M_{\rm p}/M_{\rm Jup}\lesssim 1$. 
The derived shape of the desert,  which might extend up to periods of  $5$--$10$ d,
could shed some light on the formation and evolution of close-in planets.
}
\authorrunning{Mazeh, Holczer \& Faigler }
\titlerunning{Desert of short-period Neptune-size planets}
\maketitle

\section{Introduction}
A correlation between the mass and period of the short-period Neptune-size planets was suggested early on \citep{mazeh05}, when only very few systems in this region were known. After more planets were detected, a few studies suggested the existence of a significant dearth of Neptune-mass planets with orbital periods shorter than $2$--$4$ d \citep[e.g.,][]{szabo11,benitez11,beauge13,helled16}. 
This cannot be explained by observational biases because many Neptune-mass and Neptune-radius planets with longer orbital periods were observed  (see Fig.~\ref{fig:MvsP_main} and the discussion below), and the shorter the period, the easier the detection, both by radial-velocity and by transit surveys.  
Following \citet{szabo11}, we therefore denote this region of the period-mass and period-radius planes the short-period Neptunian ``desert''. 

The existence of this desert is similar to the appearance of  the so-called brown-dwarf desert \citep[e.g.,][]{grether2006}, which suggests different formation mechanisms of planets and stellar companions with short orbital periods \citep[e.g.,][]{armitage2002}. 
Similarly,  the Neptunian desert could indicate different mechanisms of formation for hot Jupiters, with higher masses and larger
radii than the corresponding values of the desert, and short-period super-Earths, with values lower than those of the desert.    

Following \citet{szabo11}, we study here the location and shape of the desert in both the period-mass and period-radius planes using the large accumulating samples of planets available today. 
For the period-mass plane we use the set of detected planets with derived masses that appear in the exoplanet encyclopedia.\footnote{http://exoplanet.eu/catalog/} For the period-radius analysis we use two samples: the transiting planets detected by the ground-based surveys as appearing in the same encyclopedia, and the sample of \ik\ planet 
candidates\footnote{http://exoplanetarchive.ipac.caltech.edu/}, all of which have derived radii.

The set of planets with derived masses is inhomogeneous because it is based on a few surveys with different observational biases. So is the set used for the period-radius analysis, which is composed of the discoveries of two very different surveys. Nevertheless, the different sets point to a significant dearth in the period-mass and period-radius planes, as we show below. 

In Sect.~\ref{MvsP_main} we present the findings in the  period-mass plane and briefly discuss the observational biases of the different samples. In Sect.~\ref{MvsP_desert} we delineate the probable boundaries of the desert. 
Section~\ref{RvsP_main} presents the data and their observational biases for the period-radius plane, and Sect.~\ref{RvsP_desert} delineates the probable boundaries of the period-radius desert. Section~\ref{discussion} summarizes our results and briefly lists possible, sometimes contradicting, implications of the existence of the desert.

%
\section{Period-mass diagram for short-period planets}
\label{MvsP_main}

\begin{figure}[t]
\centering
{\includegraphics[width=9cm,height=9cm]{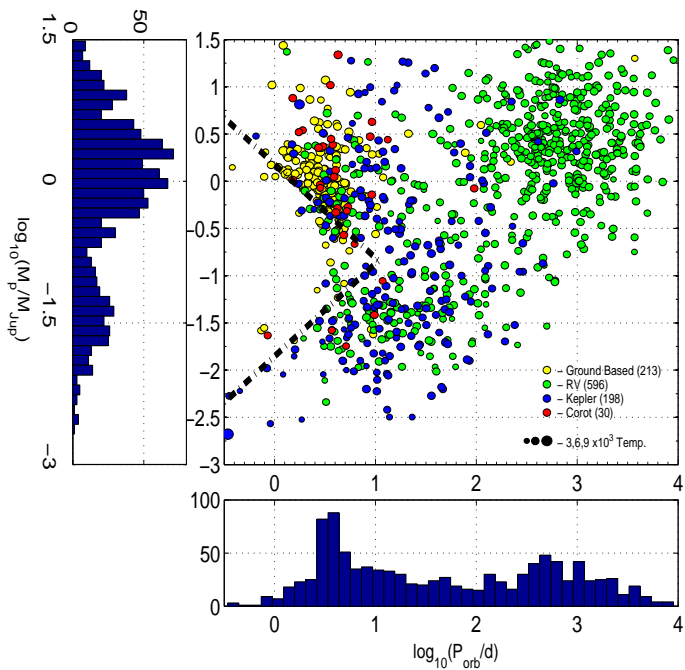}}
\caption{Planetary masses as a function of their orbital periods for planets with known masses. Planets detected by ground-based photometric searches are plotted with yellow, RV surveys with green, CoRoT detections with red, and {\it Kepler} detections with blue points. 
Point sizes represent the temperature of the parent stars.
The dash-dotted lines delineate the boundaries of the Neptunian desert, as derived below. 
}
\label{fig:MvsP_main}
\end{figure}

Figure~\ref{fig:MvsP_main} depicts  the derived masses of detected exoplanets as a function of their orbital periods taken from the exoplanet encyclopedia 
 on August 2015, with planetary masses in the range of $-3< \log_{10} (M_{\rm p}/M_{\rm Jup}) <1.5$ and orbital periods $-0.5<\log_{10}(P_{\rm orb}/{\rm d})<4$, where 
$M_{\rm p}$ is the planetary mass, $M_{\rm Jup}$ is Jupiter mass,
and $P_{\rm orb}$ is the orbital period of the planet.

The figure includes 
\begin{itemize}
\item
213 planets detected by ground-based photometric surveys  (yellow),
\item
596 planets detected by radial-velocity (RV) observations (green), 
\item
30 planets detected by CoRoT (red), and
\item
198 planets detected by \ik\ (blue).
\end{itemize}

Masses of transiting planets are obtained from the combination of photometric and RV measurements, and therefore are well constrained by the observations. Masses derived for the RV planets are only minimum masses, as their orbital inclination are not known. For this statistical analysis we have multiplied these masses by  a geometrical correction factor of $4/\pi$. 
 
Each subsample suffers from different observational biases. Transiting planets are mostly found with short orbital periods because the probability of having a transit is
\begin{equation}
{\rm P}({\rm transit})\propto a_{\rm orb}^{-1} \propto P_{\rm orb}^{-2/3} \,  , 
\end{equation}
where  $a_{\rm orb}$ is the orbital semi-major axis. The bias against longer periods is even stronger because the duty cycle of a transit, and therefore its signal-to-noise ratio (S/N), is, again, proportional to $a_{\rm orb}^{-1}$. 

The bias against detecting long-period planets  is substantially less pronounced for RV observations because the RV amplitude, $K_{\rm orb}$, a key feature in the planetary detection threshold, is  
\begin{equation}
K_{\rm orb}\propto a_{\rm orb}^{-1/2} \propto P_{\rm orb}^{-1/3}\, . 
\end{equation}
On the other hand, the samples searched for transits are much larger than those searched by RV observations. 
Therefore, in the short-period region of Fig.~\ref{fig:MvsP_main} most of the planets were detected by the transiting technique, while most of the long-period planets were detected by RV surveys. 
Another obvious observational bias is the difficulty of detecting low-mass planets, which sets the lower envelope of the period-mass graph.  

Although the subsamples in Fig.~\ref{fig:MvsP_main} are subject to different observational biases, two features are clearly visible:
the well-known paucity of Jupiter-mass planets with periods of $\sim 30$ d, and the desert of short-period Neptune-mass planets, with masses in the range of about $0.03$--$0.3\,M_{\rm Jup}$ and orbital periods below $\sim5$--$10$ d. 

 In Fig.~\ref{fig:MvsP_main} we plot two dash-dotted lines that enclose the Neptunian desert. The slope and location of each of the two lines were chosen to produce the best contrast between the desert and its surroundings, as detailed in the next section.
We note that although the two lines intersect at  $\sim 10$ d, it is not clear that the desert extends up to this period because the picture is not clear at the range of $\sim5$--$10$ d. The two lines delineate the boundaries of the desert at shorter periods.
 
\section{Neptunian desert at the period-mass plane}
\label{MvsP_desert}
 To facilitate the discussion, we transformed the planetary mass and orbital period into the variables
\begin{equation}
 \mathcal{M}= \log_{10} (M_{\rm p}/M_{\rm Jup})
\ \  {\rm and} \ \ 
\mathcal{P}=\log_{10} (P_{\rm orb}/{\rm d}) \, ,
\end{equation}
as used in Fig.~\ref{fig:MvsP_main}. In what follows we present the analyses of the upper and lower boundaries of the desert separately.

\subsection{Upper boundary of the period-mass desert}    %

We assumed the upper boundary of the desert to be characterized by a line in the ($\mathcal{P},\mathcal{M}$) plane
\begin{equation}
 \mathcal{M}= a\mathcal{P}+b\, ,
\end{equation}
so that the planetary occurrence is low below the line and high above it, as plotted in Fig.~\ref{fig:MvsP_main}. 
Furthermore, we assumed that the planetary density in the period-mass plane is a function of the distance from this line alone.
To find the best line representing the boundary, we applied two statistical approaches.

In the first approach we divided the ($\mathcal{P},\mathcal{M}$) plane into stripes parallel to an assumed boundary line and derived the planet density in each stripe. We then changed the slope, $a$, and location, $b$, of the line and searched for the values that enhanced the density contrast between stripes below and above the boundary line.

The second approach used the maximum likelihood technique to find a probability distribution function,
$\mathcal{F}_{\rm PDF}(\mathcal{P, M})$, with a low probability below the desert boundary and high above it. Two crucial parameters of the probability were the two parameters of the line, $a$ and $b$, which carried the information about the desert boundary. We tried different values of $a$ and $b$ to find the values that maximized the likelihood of the given sample. 
 
In both approaches we focused on a circle in the ($\mathcal{P},\mathcal{M}$) plane, where the density contrast of the desert was most pronounced, 
as plotted in Fig.~\ref{fig:MvsP_upper_stripes}. 
The circle, with a center  at $(\mathcal{P, M})=(0.3,-0.2)$ and a radius of $0.4$, included 152 planets. We used a circle to minimize biases that might arise from choosing a rectangle, which inherently has a preferred direction. A few different circles did not change our results.

%
\subsubsection{Stripes to maximize the density contrast}   %
\label{MvsP_stripes}

\begin{figure}[t]
\centering
{\includegraphics[width=6.5cm,height=6.5cm]{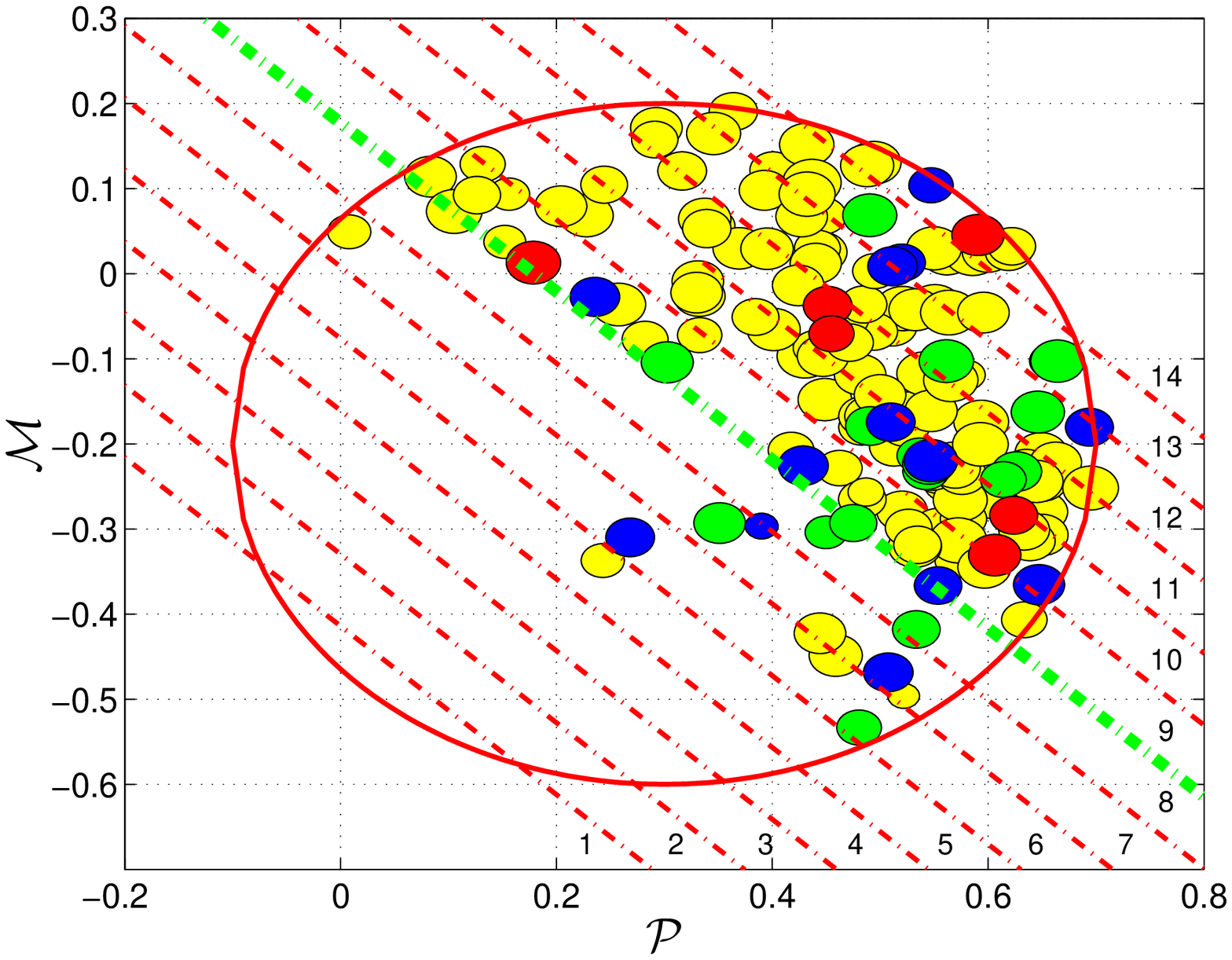}
\includegraphics[width=6.5cm,height=6.5cm]{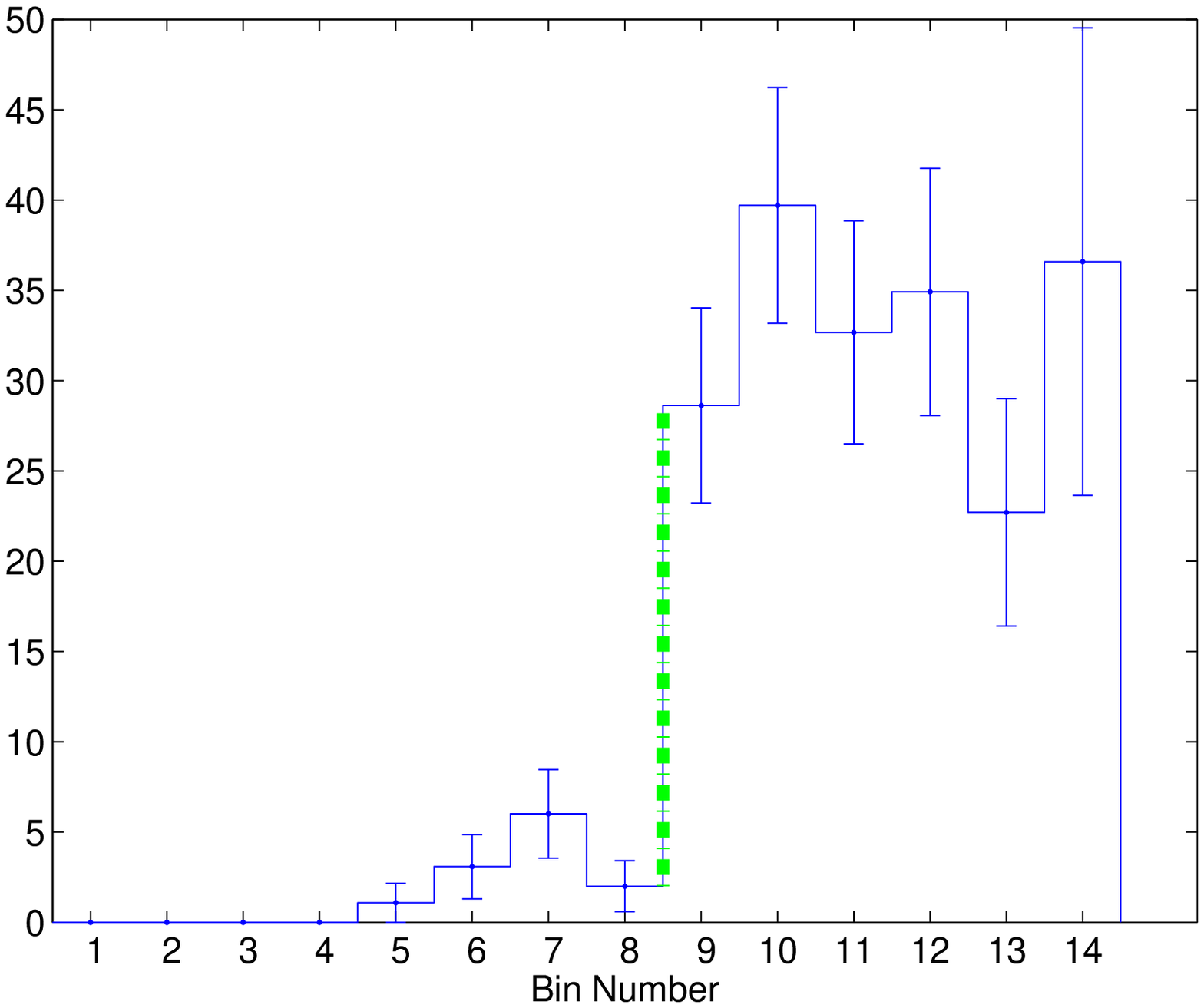}
}
\caption{Stripe technique for deriving the desert boundary.
Upper panel: The circle over which the analysis was performed, divided this time into 14 numbered stripes. The slope of the stripes shown yielded the highest contrast.
Lower panel: Histogram plot of the number of planets in each stripe, normalized to its area inside the circle. }
\label{fig:MvsP_upper_stripes}
\end{figure}

Figure~\ref{fig:MvsP_upper_stripes} presents our first approach. We divided the circle into parallel stripes and derived a histogram that presented the number of planets found in each stripe. This was done after correcting for the different areas of the stripes, adopting a unity area for the largest stripe.
The errors of each of the histogram bins were the square root of the number of planets in each stripe, also corrected for the stripe area.
 
The slope of the stripes and their exact location were searched for the best possible contrast in planet number, defined for any two adjacent stripes $i$ and $i+1$ with corrected 
number of planets $n_i$ and $n_{i+1}$, respectively, as
\begin{equation}
 \Re=\frac{n_{i+1}-n_i} {\sqrt{n_{i+1}^2+n_i^2}}\, .
\end{equation}

The search yielded a sharp boundary, plotted in Fig.~\ref{fig:MvsP_upper_stripes}, with a contrast of $\Re=4.8$. This was found for a stripe width of 0.06 and a boundary at
\begin{equation}
\mathcal{M}=-0.99\,\mathcal{P}+0.18\, .
\end{equation}
%

\subsubsection{Maximum likelihood with a modified Fermi function}   %
\label{MPmaxlike}                                                         %

The second algorithm assumed  a modified Fermi function for the density distribution of the planets. The density was a function of the distance, $d=d(a,b)$, from the  
$\mathcal{M}= a\mathcal{P}+b$ line:
\begin{equation}
\mathcal{F}_{\rm PDF}(d(a,b); \delta, \Delta)=
\mathcal{A} \{ \frac{1}{1 + exp( -\frac{d}{\delta})}+\Delta \}\, ,
\end{equation}
where $d$ was positive above the line and negative below it, 
$\delta$ was the transition width, and $\Delta$ was the low density in the desert. 
The constant $\mathcal{A}$ was defined so that the 2D integral over the circle area, using the four parameters, $a, b, \delta$ and $\Delta$, equaled unity. 

For each set of parameters, we derived the likelihood of the given sample of planets included in the circle of Fig.~\ref{fig:MvsP_upper_stripes},
 \{$(\mathcal{P}_i, \mathcal{M}_i);i=1,N$\}, by
\begin{equation}
\mathcal{L}= \prod_{i=1}^N \mathcal{F}_{\rm PDF}(\mathcal{P}_i,\mathcal{M}_i;a, b, \delta, \Delta) \, ,
\end{equation}
 and ran an MCMC routine with 250 walkers and $10^6$ steps \citep[e.g.,][]{mackey13} to find the parameter values that maximized the sample likelihood and their uncertainties.
 
The best line we found was
\begin{equation}
\mathcal{M}=-(1.14\pm0.14)  \mathcal{P}+(0.230\pm0.045)  \, ,
\end{equation}
while $\delta=0.0214 \pm 0.0036 $ and $\Delta=0.048 \pm0.016$.
 

These parameters are within about $1\sigma$ of the values found by the stripe approach. 

\subsection{A lower boundary of the period-mass desert?}

\begin{figure}[t]
\centering
{\includegraphics[width=6.5cm,height=6.5cm]{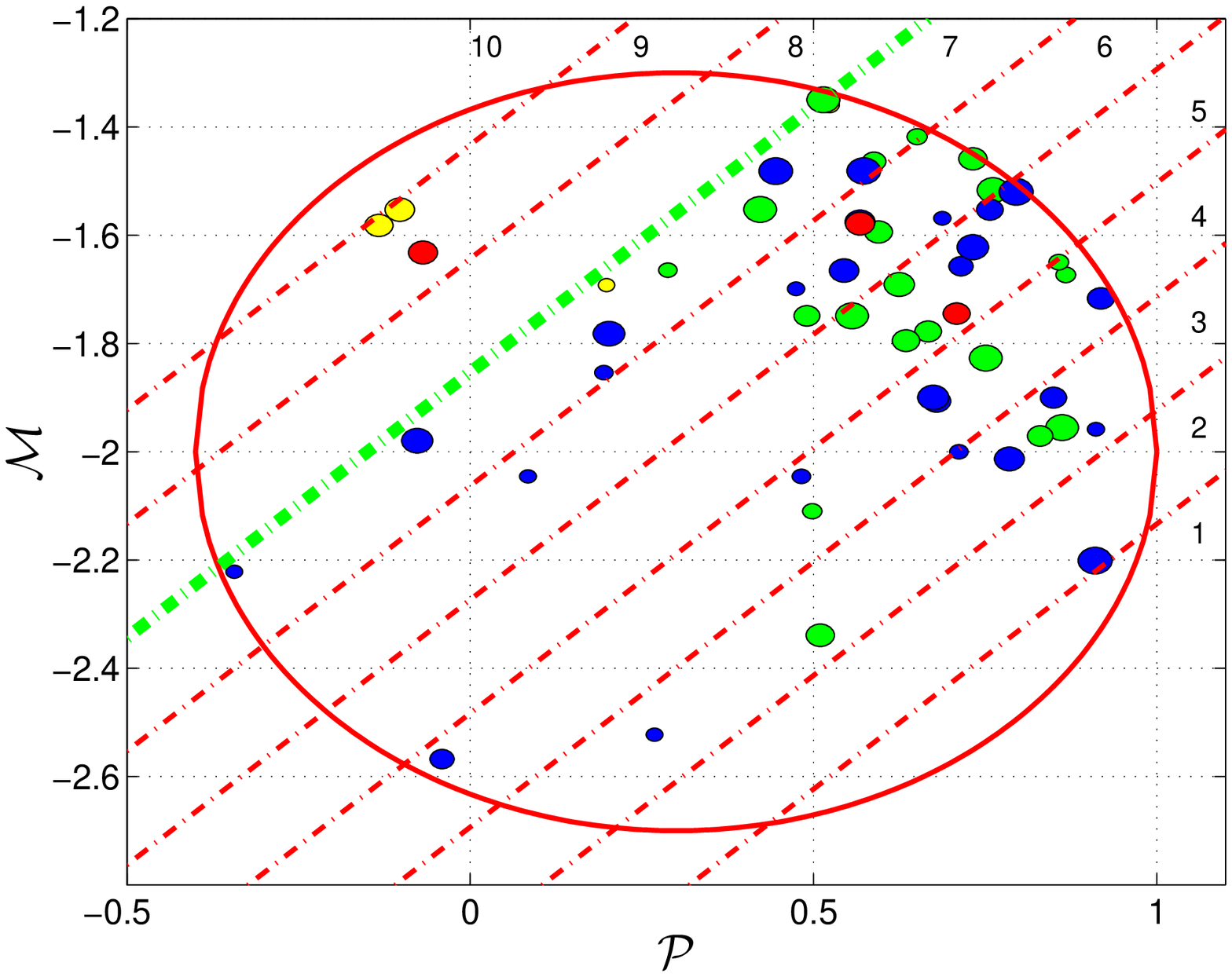}
\includegraphics[width=6.5cm,height=6.5cm]{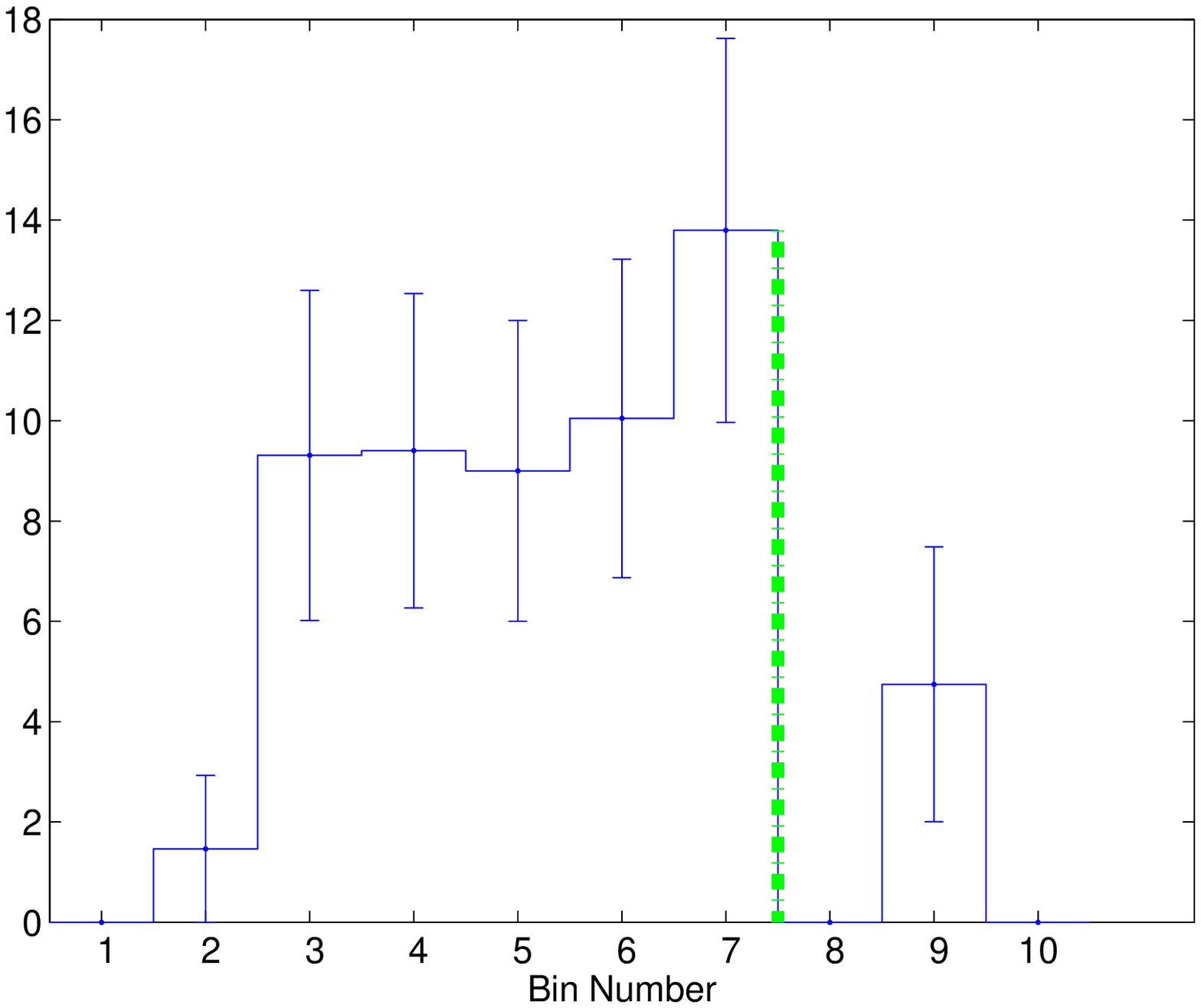}
}
\caption{
Upper panel: Planetary masses as a function of planetary orbital periods 
for planets inside the circle, divided into ten stripes. 
Lower panel: Histogram plot of the number of planets in each stripe, normalized to the stripe area inside the circle.
}
\label{fig:MvsP_lower_stripes}
\end{figure}

The lower boundary of the period-mass desert is quite blurred, probably because the number density of planets in that part of the period-mass plane is low, which is due to the observational biases and the difficulties of obtaining reliable low masses. Therefore, we were unable to apply the MCMC approach, and only the stripe technique yielded reasonable results.
We again focused on a circle in the ($\mathcal{P},\mathcal{M}$) plane, as plotted in Fig.~\ref{fig:MvsP_lower_stripes}. 
The circle, with a center at
$(\mathcal{P, M})=(0.3,-2.0)$ and a radius of $0.7$,
included 53 planets.

Although we only had a very small number of planets, most of them in a small section of the circle, and even though three planets were located in the middle of the desert, the results suggested a clear contrast, with a line at
%
\begin{equation}
\mathcal{M}=0.98\mathcal{P}-1.85\, ,
\end{equation}
as shown in Fig.~\ref{fig:MvsP_lower_stripes}.

We therefore suggest that the two boundaries of the period-mass desert might have opposite slopes of $\sim \pm1$ in the log-log plane.  The two boundaries are plotted in Fig.~1. 

We note that the desert is not completely dry. There are planets both below the upper and above the lower boundaries of the desert. Nevertheless, Fig.~1 suggests a clear contrast of planetary occurrence between the desert and its surroundings in the range that extends up to 5--10 d.

%
\section{Period-radius diagram for short-period planets}        %
\label{RvsP_main}
 
\begin{figure}[t]
\centering
{\includegraphics[width=9.5cm,height=9.5cm]{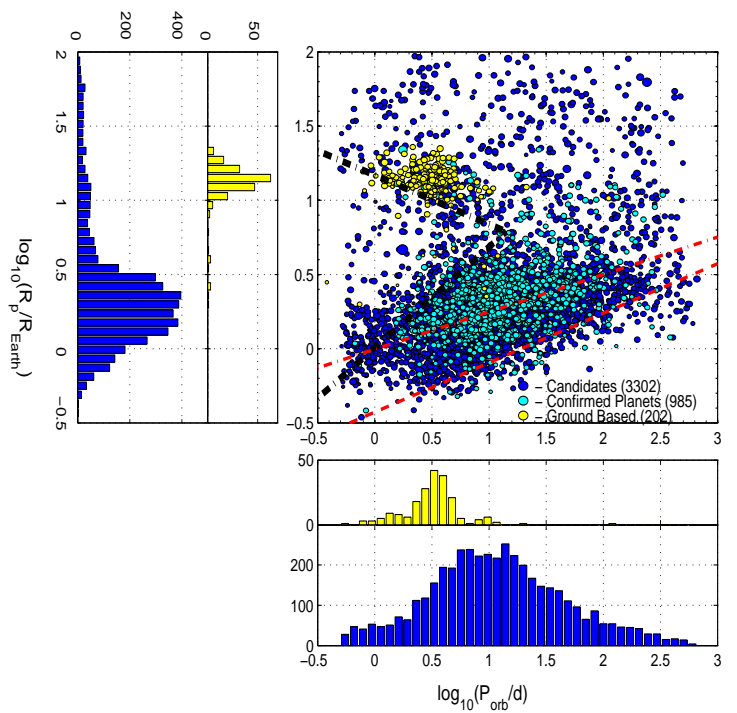}}
\caption{
Planetary radii as a function of planetary orbital periods for the ground-based transit searches (202 yellow points), {\it Kepler} planet candidates (3302 blue points), and \ik\ confirmed planets  (985 cyan points). The two dash-dotted black lines are our best-estimate desert boundaries, as derived in Sect.~\ref{RvsP_desert}.
The upper (dash-dotted) red line shows the center of the ridge and the lower (dashed) red line the observational detection bias. The histograms below and to the left of the main panel separately show the \ik\ detections, candidates and confirmed planets alike, and the ground-based detections.
}
\label{fig:RvsP_main}
\end{figure}

Figure~\ref{fig:RvsP_main} depicts the transiting planets with derived radii.  
The figure represents two sets: the planets discovered by ground-based searches for transits taken from the exoplanet encyclopedia,
and the \ik\ sample as it appeared on the \ik\ site. 
The figure shows planets with radii in the range of 
$-0,5< \log_{10} (R_{\rm p}/R_{\rm Earth}) <2$ and orbital periods of $-0.5<\log_{10} (P_{\rm orb}/{\rm d})<3$, where 
$R_{\rm p}$ is the planetary radius and $R_{\rm Earth}$ is the
Earth radius.

 To facilitate the discussion, we used the variables
\begin{equation}
 \mathcal{R}= \log_{10} (R_{\rm p}/R_{\rm Earth})
\ \  {\rm and} \ \ 
\mathcal{P}=\log_{10} (P_{\rm orb}/{\rm d}) \, ,
\end{equation}
as used in Fig.~\ref{fig:RvsP_main}.

The figure includes 
\begin{itemize}
\item
202 planets detected by ground-based transit surveys  (yellow),
\item
985 confirmed \ik\  planets (cyan), and 
\item
3302 \ik\ candidates (blue). 
\end{itemize}

The two sets are very different. On one hand, the ground-based searches surveyed a large number of stars, on the order of $10^{7}$, while \ik\ observed only $\sim2 \times 10^5$ objects. On the other hand, the ground-based searches were only able to detect planets with radii larger than $\sim 10\, R_{\rm Earth}$, while the \ik\ sensitivity reached $\sim1\, R_{\rm Earth}$ and even lower, depending on the orbital period. 
The \ik\ results show that the frequency of short-period planets with radii larger than  $3\, R_{\rm Earth}$ is quite low relative to those with smaller radii, as seen in  the radius histogram at the side of Fig.~\ref{fig:RvsP_main}. 
Therefore, the \ik\ sample was too small to explore the statistical features of the large planets, while the ground-based searches were blind to the smaller planets.
Thus, only the combination of the two sets of planets can outline the shape of the Neptunian desert, as discussed below.  We therefore opted to derive the upper boundary of the desert solely from the ground-based findings, and used only the \ik\ detections to derive the lower boundary.    

One main feature apparent in Fig.~\ref{fig:RvsP_main} is  a dense region of \ik\ detected planets, extending from a small, short-period region, at  $(\mathcal{P},\mathcal{R})\sim(0,0)$  to $\sim(1.75,0.5)$
 \citep[][]{helled16}. 
To derive the best slope of this ridge, we applied the MCMC technique, as presented in Sect. \ref{MPmaxlike}, to the ($\mathcal{P}$, $\mathcal{R}$) plane. This time, we used a normal distribution function, which fit the data better than the Fermi function. The density was assumed to be a function of the distance from the ridge line alone. 
The best line for the {\it \textup{ridge}} was found to be
\begin{equation}
\mathcal{R}= (0.25\pm0.024) \mathcal{P} - (0.003\pm0.025) \, ,
\end{equation}
with  a broadening width of $0.22\pm0.01$. 
The ridge best-fit line is plotted as a dash-dotted red line in Fig.~\ref{fig:RvsP_main}.


Below the ridge lies a parameter region that is poorly represented by the \ik\ set because of the detection threshold  \citep[][]{helled16}
that is determined by the S/N of a transit, which  for a given stellar radius is proportional to $R_{\rm p}^2/a_{\rm orb}$ (see Sect. 2), or $R_{\rm p}^2/P_{\rm orb}^{2/3}$.
Therefore, a constant S/N is along a line for which $R_{\rm p} \propto P_{\rm orb}^{1/3}$. The lower dashed red line in Fig.~\ref{fig:RvsP_main} denotes such a possible threshold. Below this lower line, the S/N is too low to allow for frequent detections. The few planets below this line are probably transiting small or quiet parent stars. We found the exact position of the line by another MCMC run, this time assuming a slope of one-third in the 
 $(\mathcal{P},\mathcal{R})$ plane.

Above the ridge, at larger radii, the planet density gradually
decreases.
This is not an observational bias, as larger planets are easier to detect. Above the gradual decrease, a clear desert emerges.  
We therefore added two dash-dotted black lines that are assumed to border the Neptunian desert, as discussed below. The slope and location of each of the two lines were chosen to produce the highest contrast between the two sides of that border line.
Obviously, the desert is not completely dry. There are planets both below the upper and above the lower boundaries of the desert. Nevertheless, the occurrence of planets in the desert is substantially lower than outside its borders.

%
\section{Neptunian desert at the period-radius plane}    %
\label{RvsP_desert}

\subsection{Upper boundary of the period-radius desert}    %

\begin{figure}[t]
\centering
{\includegraphics[width=6.5cm,height=6.5cm]{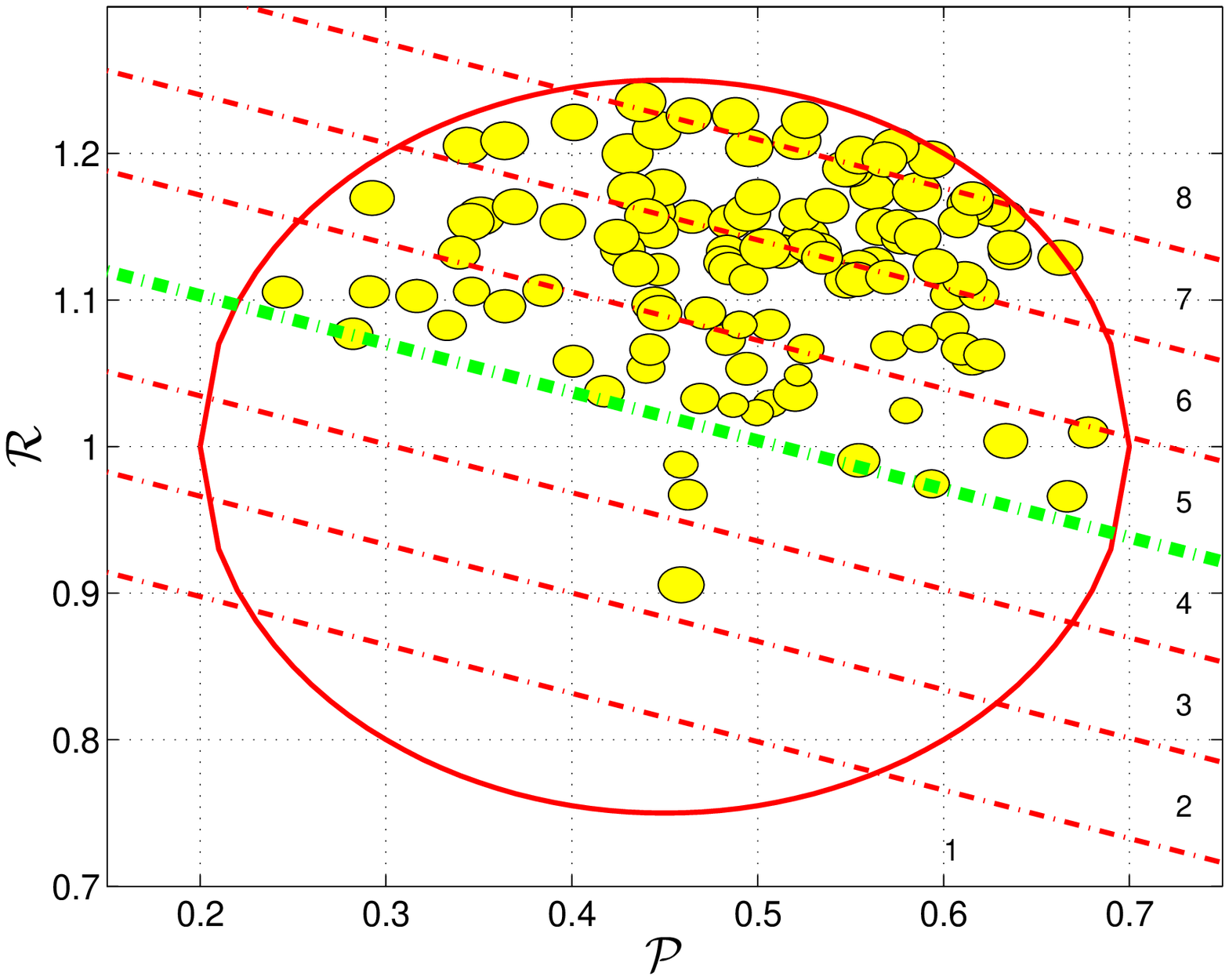}
\includegraphics[width=6.5cm,height=6.5cm]{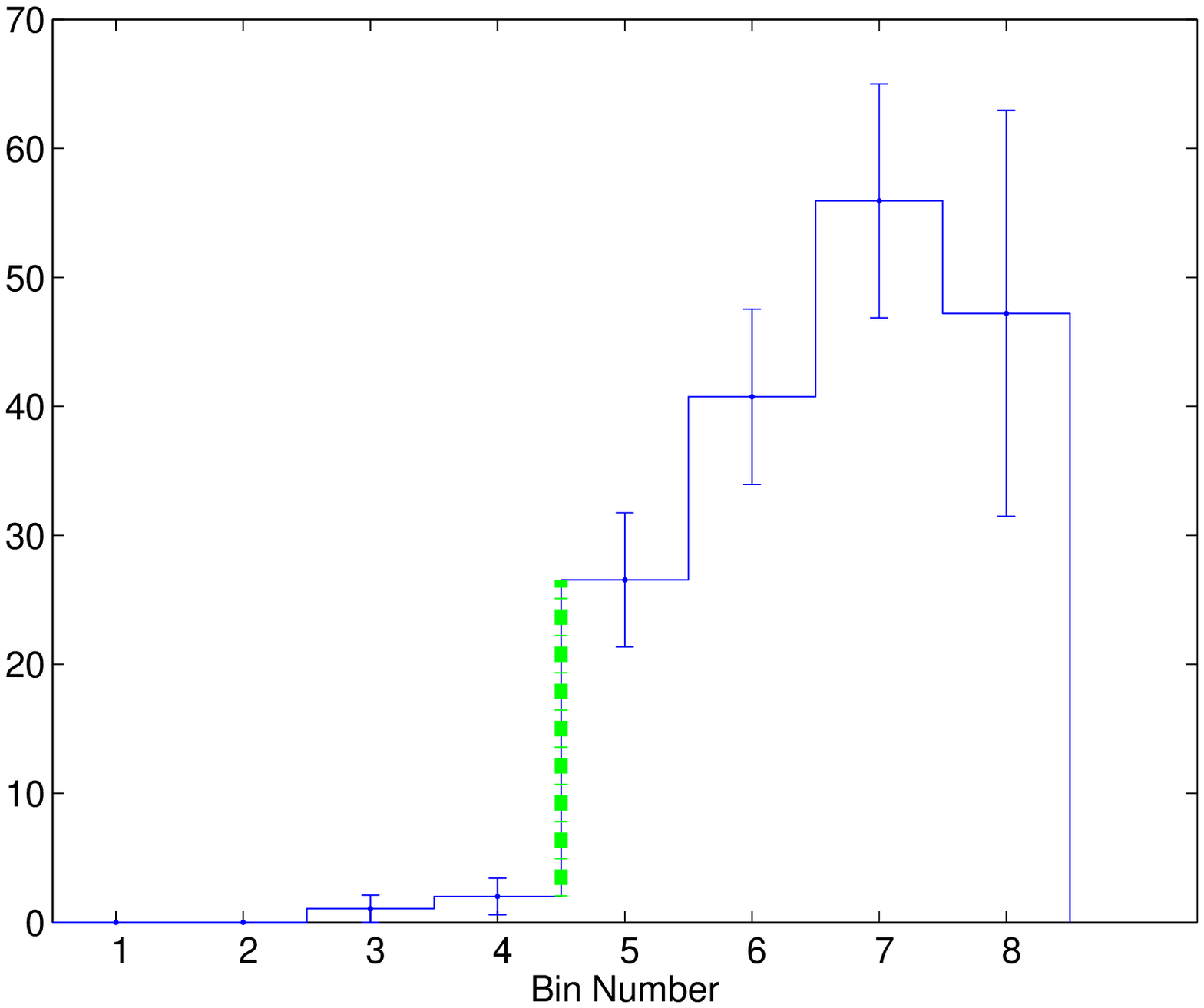}
}
\caption{Stripe technique for deriving the upper boundary of the period-radius desert.
Only ground-based detections were used. 
Upper panel: The circle in which the analysis was performed, as in Fig.~\ref{fig:MvsP_upper_stripes}, divided this time into eight numbered stripes. The slope of the stripes shown yielded the highest contrast.
Lower panel: Histogram plot of the number of planets in each stripe, normalized to its area inside the circle. 
}
\label{fig:RvsP_upper_stripes}
\end{figure}

To determine the upper boundary of the period-radius desert, we assumed, as in the period-mass analysis, that it was characterized by a line in the ($\mathcal{P},\mathcal{R}$) plane
\begin{equation}
 \mathcal{R}= a_{R}\mathcal{P}+b_{R}\, .
\end{equation}
We assumed the planet density to be low below the line and high above it and that it is a function of the distance from this line alone.
To determine the best-fit line, we applied the same two statistical approaches.
In both approaches we focused on a circle in the period-radius plane, with a center at $(\mathcal{P, R})=(0.45,1)$ and a radius of $0.25$, which included 112 planets detected by the ground-based surveys, as depicted in Fig.~\ref{fig:RvsP_upper_stripes}.

The stripe analysis found the highest contrast, $\Re=4.5$, for a stripe width of 0.06 and a boundary at
\begin{equation}
\mathcal{R}=-0.33\,\mathcal{P}+1.17\, .
\end{equation}
The best-fit boundary found by the likelihood approach with the modified Fermi density function was
\begin{equation}
\mathcal{R}=-(0.31\pm0.12)  \mathcal{P}  +  (1.19\pm0.06)  \, ,
\end{equation}
while $\delta=0.021 \pm 0.003 $ and $\Delta=0.03 \pm0.03$.

These parameters are within less than $1\sigma$ of the values found by the stripe approach. We therefore adopted the values found by the stripe approach and plot its derived line in Fig.~\ref{fig:RvsP_main}.

\subsection{Lower boundary of the period-radius desert}

\begin{figure}[t]
\centering
{\includegraphics[width=6.5cm,height=6.5cm]{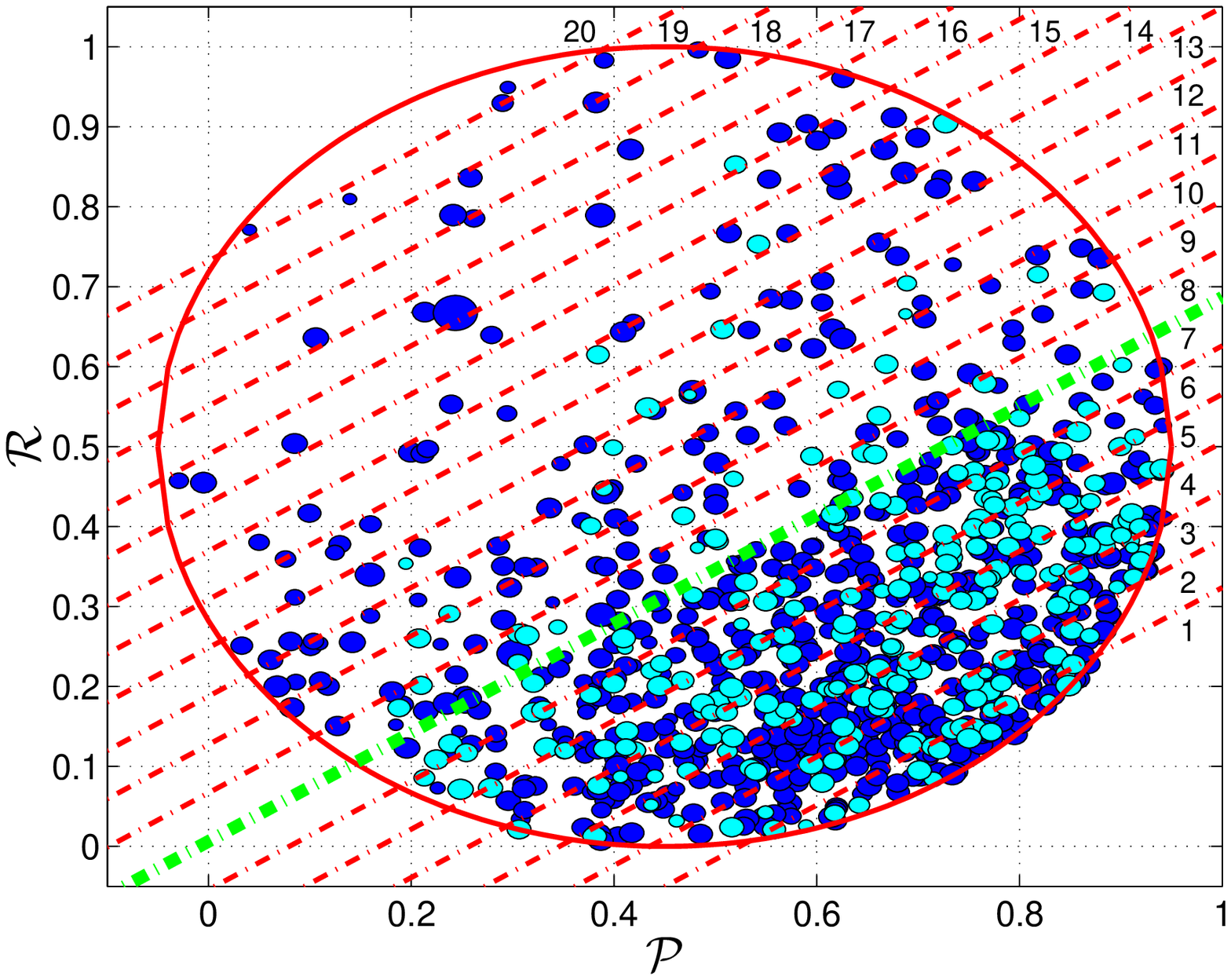}
\includegraphics[width=6.5cm,height=6.5cm]{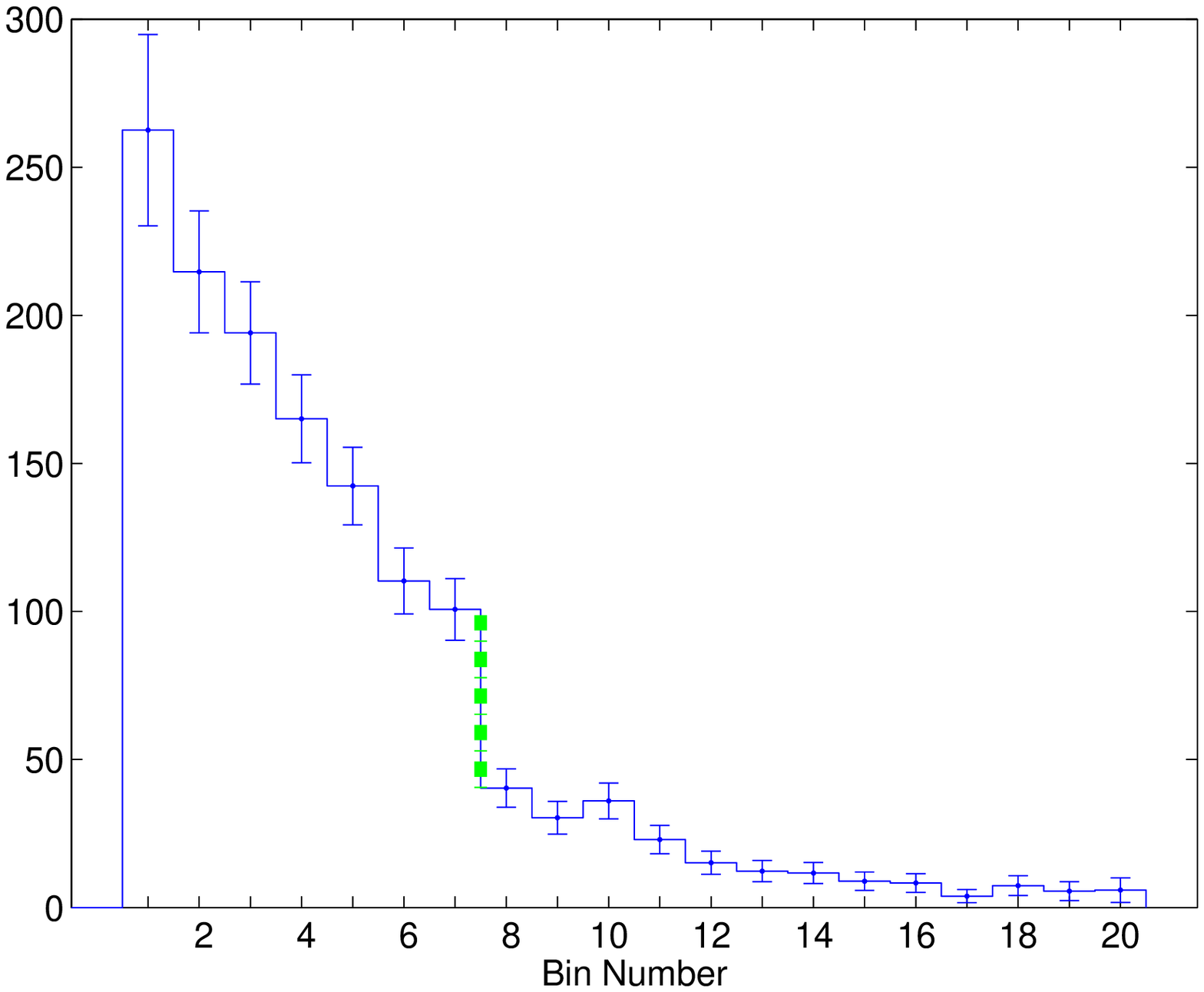}
}
\caption{Stripe technique for deriving the boundary of the period-radius desert.
Upper panel: The circle in which the analysis was performed, as in the previous figure, divided this time into 20 numbered stripes. The slope of the stripes shown yielded the highest contrast.
Lower panel: Histogram plot of the number of planets in each stripe, normalized to its area inside the circle. 
}
\label{fig:RvsP_lower_stripes}
\end{figure}

To determine the lower boundary of the desert, we again applied our two approaches, focusing on a circle in the ($\mathcal{P},\mathcal{R}$) plane, as plotted in 
Fig.~\ref{fig:RvsP_lower_stripes}, using this time only the \ik\ detections.
The circle, with a center at
$(\mathcal{P, R})=(0.45,0.5)$ and a radius of $0.5$,
included 929 candidate and confirmed planets.

The stripe analysis (see Fig.~\ref{fig:RvsP_lower_stripes}) yielded a contrast of $\Re=4.9$ for a stripe width of 0.05 and a transition at
\begin{equation}
\mathcal{R}=0.68\,\mathcal{P}\, .
\end{equation}
Figures~\ref{fig:RvsP_main} and \ref{fig:RvsP_lower_stripes} suggest that  the appearance of the desert is convolved with the gradual decay of the planet density above the ridge. Nevertheless, the lower edge of the desert can still be clearly seen. 

The best-fit  boundary found by the likelihood approach with the modified Fermi density function was
\begin{equation}
\mathcal{R}=(0.67\pm0.06)  \mathcal{P}-(0.01\pm0.04)  \, ,
\end{equation}
while $\delta=0.026 \pm 0.003 $ and $\Delta=0.10 \pm0.02$. The relatively high value of $\Delta$ indicates that the desert is not as dry in the period-radius plane. 
These parameters are within less than $1\sigma$ of the values found by the stripe approach. We therefore adopted the values found by the stripe approach and plot its derived line in Fig.~\ref{fig:RvsP_main}.

The two derived lines delineate the boundaries of the desert at the period-radius plane, a region that extends up to $\sim 5$ d.

\section{Discussion}    %
\label{discussion}       %

The analysis presented here 
establishes the existence of a short-period Neptunian desert. Although the sets of planets with derived masses and radii we used were subject to different observational biases, the desert appears both in the planetary period-mass and period-radius planes. 
The desert might indicate two populations: Jovian and super-Earth planets. The super-Earths were mostly detected by the \ik\ mission, while the Jovian planets, which are much less frequent, were
detected mainly by the ground-based transit surveys.  

The upper boundary of the desert is clearly seen in both planes. 
The lower boundary is less significant. 
It is barely seen in the period-mass plane, probably as a result
of observational biases, but can be better seen in the period-radius plane, although admittedly not as well as the upper boundary. 

The upper bounds of the desert, at the range of
$0.1\lesssim M_{\rm p}/M_{\rm Jup}\lesssim1$, as seen in the period-mass and period-radius planes,  are at about 
\begin{equation}
M_{\rm p}/M_{\rm Jup}\simeq(1.7\pm0.2)(P_{\rm orb}/{\rm d})^{-1.14\pm0.14} 
\end{equation}
or
\begin{equation}
R_{\rm p}/R_{\rm Jup}\simeq(1.4\pm0.3)(P_{\rm orb}/{\rm d})^{0.31\pm0.12} 
 \, .
\end{equation}
If these two boundaries represent the same edge, then along this line the mass-radius relation is  
\begin{equation}
R_{\rm p}/R_{\rm Jup}\simeq (1.2\pm0.3)(M_{\rm p}/M_{\rm Jup})^{0.27\pm0.11}
 \, .
\end{equation}
For this range of masses, $0.1\lesssim M_{\rm p}/M_{\rm Jup}\lesssim1$, 
the mass-radius relation is similar to the one used by \citet{matsakos16}, which was based on the relation reported by \citet{weiss13}.

The upper and lower boundaries in both the period-mass and period-radius planes intersect at 
 $\sim 10$ d. However, as emphasized above,  it is not clear that the desert extends up to this period.  The boundaries we derived, with all the reservations that were pointed out in the analysis, delineate the boundaries of the desert at shorter periods, below $\sim 5$ d.

The goal of this paper is not to discuss the astrophysical processes that cause the desert. Nevertheless, we list below a few possible, sometimes conflicting, interpretations. We start with the approach that separates the hot Jupiters, with the upper boundary of the desert, from the super-Earths that determine the lower boundary. 

In the framework of planetary migration, we mention two processes that might account for the upper boundary of the desert.
\begin{itemize}
\item
During the migration of the planets, the upper boundary acted like a death line beyond which planets could not exist.
Planets that moved horizontally  in Fig.~\ref{fig:MvsP_main} or \ref{fig:RvsP_main} under the migration processes and crossed their death line were doomed because they lost a major part of their masses, either 
due to the stellar insolation \citep[e.g.,][]{lopez14, tian15} or because of Roche-lobe overflow (\citet{kurokawa14, matsakos16}, but see also \citet{mustill15,inamdar15,mordasini15} and \citet{hansen15}). Consequently, they moved down in the period-mass or the period-radius plane into the ridge below the desert. Only planets that stopped their migration before they crossed their death line have survived with their high masses. The death line depends on the planetary mass because more massive planets can migrate closer to their host star without being stripped.
\item
The migration driven by the interaction with the disk stopped near the upper edge of the desert  because at that point the disk was not dense enough to continue pushing the planets inward.  
This might be due to the central hole in the accretion disk, induced by some interaction with the star, the stellar magnetic field, for example. If this were the case, the inner radius of the disk might be related to its mass and therefore to the planetary mass. The more massive the disk, and hence the planet, the smaller its central hole.
\end{itemize}  
Obviously, to model the short-period desert and its upper boundary, any migration scenario needs to be worked out with detailed numerical considerations and simulations. 

To account for the lower boundary of the desert and for the existence of the super-Earth ridge, we mention the 
recent works that  suggested {\it \textup{in situ}} models for the formation of short-period super-Earths
 (\citet{lee16}; see also \citet{hansen12,chiang13} and \citet{lee14}). These are based on the widely used core-accretion planet formation models
 \citep[e.g.,][]{pollack96,helled14, robinson09,rogers11}, which assume three stages of planetary formation. The first stage is the fast buildup of a solid planet core on the order of $10 M_{\rm Earth}$. The second stage involves a slow and long hydrostatic accumulation of gas through Kelvin-Helmholz contraction, until the envelope mass is similar to that of the core. The third stage is then triggered as a fast runaway gas accretion that can produce a giant planet. The formation process may end at any of these stages with the depletion of the protoplanetary disk at the vicinity of the forming planet. The final mass of the planet thus depends, among other factors, on the relative timescales of the disk evolution and the planet formation process.

The lower boundary of the desert and the super-Earth ridge, with planetary masses monotonically increasing with orbital period, can be explained by a variety of effects that determine the final planetary mass and vary as a function of the separation from the central star. 
At larger separation we might find, 
for example, a higher volume of the Hill sphere of the forming planet, a longer orbital path, and a higher dust-to-gas ratio. These effects increase the core mass at the end of the first stage, thus increasing the final planetary mass of most planets at larger separations, which explains the monotonicity of the ridge and the lower boundary of the desert.
 
Some recent works considered hot Jupiters and super-Earths to
be formed by the same mechanism, and therefore may interpret the desert as a whole. 
One such work \citep{matsakos16} suggested an ingenious model in which short-period 
Jupiters and super-Earths both moved toward their parent stars by high-eccentricity migration and then were tidally circularized in the vicinity of their Roche limit, which is the reason for the observed desert. This model explains the slopes of the upper and lower boundaries of the desert as a consequence of the different mass–radius relation for small and large planets. 
 
Another suggestion that considered a common formation mechanism for super-Earths and hot Jupiters has been made by \citet{batygin15}, who argued for {\it \textup{in situ}} formation of hot Jupiters. According to this approach, the planetary mass increases to $\sim 15 M_{\rm Earth}$ for most cases during the first two formation stages,
and it rapidly crosses the desert mass range of $\sim 20$--$100 M_{\rm Earth}$ during the fast, very short-lived third stage. This means that producing a planet with a mass within this range requires an unlikely detailed fine-tuning of the formation and the disk evolution processes.
Furthermore, the paucity of hot Jupiters relative to super-Earths indicates that the disk lifetime is usually shorter than the planet formation timescale. This explains why only few planets reach the third stage, go through the gas accretion runaway, and become hot Jupiters.
However, we suggest that this interpretation of the desert in less likely because we cannot conceive how this approach might account for the clearly significant upper boundary of the desert.  
   
To summarize, we suggest that the ridge of the super-Earths and the short-period Neptunian desert in particular, at
$\sim0.03$--$0.3\, M_{Jup}$ or
$\sim3$--$10\, R_{Earth}$,
might indicate the existence of two distinctive short-period planet types, Jovians and super-Earths, suggesting two different formation processes for the two populations. 
In this sense, the Neptunian desert might be similar to the short-period brown-dwarf desert
\citep[e.g.,][]{grether2006, cheetham15, bouchy15}, in the range of 
$\sim20$--$80\, M_{Jup}$, which distinguishes between planets and stellar companions 
 \citep[e.g.,][]{armitage2002, ma14, brandt14, thies15}.
The two deserts, which are located apart by a factor of 300 in mass, indicate three distinctive populations of short-period companions: stellar companions, Jovian planets, and super-Earth planets. 
It would be of interest to determine whether the boundaries of the Neptunian desert depend on the stellar temperature and metallicity, as seems to be the case for the brown-dwarf desert \citep{deleuil08}.

\begin{acknowledgements}
We thank Flavien Kiefer, who worked with us at the beginning of the project. The invaluable advice of Shay Zucker about the analysis and the careful reading of the manuscript and very thoughtful comments of Arieh Konigl, Titos Matsakos, and Daniel C. Fabrycky are much appreciated.  
This research has received funding from the European Community's Seventh Framework Programme (FP7/2007-2013) under grant-agreement numbers 291352 (ERC).
\end{acknowledgements}


\end{document}